\documentclass{MolSpaLab}
\usepackage{graphicx}
\begin{document}
\TitreGlobal{Molecules in Space \& Laboratory}
\title{The coupling of dynamics and molecular chemistry in galaxies}
\author{\FirstName F. \LastName Combes}
\address{Observatoire de Paris, LERMA, 61 Av. de l'Observatoire, F-75014, Paris}
\runningtitle{Dynamics and chemistry in galaxies}
\setcounter{page}{1}

\maketitle
\begin{abstract}
While the best tracer of the molecular component and its dynamics in galaxies
is the CO molecule, which excitation is revealed by its isotopic and (2-1)/(1-0) ratios,
the denser gas is revealed by molecules such as HCN, HNC, HCO$^+$ or CN, which are
now widely used to probe star formation regions, or to quantify the impact of
the nuclear activity on the interstellar medium.
 This paper reviews recent observations in nearby galaxies, where these molecular
line ratios  serve as diagnostic tools of the physical conditions of the gas
and also of its chemical properties.  Those differ significantly according to
the proximity of an AGN or of a starburst. The origin of the differences
is not yet well known and could be due to different densities, temperatures,
chemical abundances or non-collisional excitation of the gas
(e.g. Aalto et al 2007, Krips et al 2007).

HCN or HNC line enhancements can be caused not only by higher gas densities/temperatures,
but also UV/X-ray radiation, and global IR pumping.
The chemistry can be dominated by PDR regions near a starburst, or
X-ray dominated in a molecular torus surrounding an AGN (XDR regions).
The molecular line ratios expected in those regions vary according to the
different models (Meijerink et al. 2007).
\end{abstract}
%

\bigskip

In the last decade, a large progress has
been made in our knowledge of various molecules
other than CO in external galaxies. 
A wide molecular survey has been carried out in the nearby starburst 
galaxy NGC~253 (Martin et al 2006), and 
35 (4 tentative) new species have been detected,
with in addition 13 (+2) isotopic substitutes
(cf Table 1).

\begin{table}
\caption{Extragalactic Molecules}
\begin{tabular}{cccccc}
\hline
2 atoms &  3 atoms   & 4 atoms  &   5 atoms   &   6 atoms &   7 atoms \\
\hline
H$_2$      &      H$_2$O   &     H$_2$CO &     c-C$_3$H$_2$  &    CH$_3$OH  &    CH$_3$CCH \\
OH      &      HCN   &     H$_2$CS &     HC$_3$N    &    CH$_3$CN  &           \\
CO      &      HNC   &     NH$_3$  &     CH$_2$NH   &           &           \\
CH      &      HCO   &     HNCO &     NH$_2$CN   &           &           \\
CS      &      HCO$^+$  &     C$_3$H  &             &           &           \\
CH$^+$     &      H$_2$S   &     HOCO$^+$&             &           &           \\
CO$^+$     &      SO$_2$   &          &             &           &           \\
NO      &      C$_2$H   &         &             &           &           \\
CN      &      HOC$^+$  &         &             &           &           \\
NS      &      C$_2$S   &         &             &           &           \\
SiO     &      N$_2$H$^+$  &         &             &           &           \\
SO      &      OCS   &         &             &           &           \\
LiH     &      H$_3$$^+$   &         &             &           &           \\
\hline
\end{tabular}
\end{table}

\section{Shocks and Chemistry in Galaxies}

The silicon monoxide molecule (SiO) is well established
as a good tracer of high temperatures and/or shock chemistry
in molecular clouds. For shock velocities larger than 40 km/s,
grain destruction becomes important, and Si and SiO can be
released to the gas phase.
SiO abundance is enhanced in star-forming regions, 
supernovae remnants, but not in clouds or PDR
(photon dominated regions).

SiO emission is intense in the Seyfert 2 prototype NGC1068,
and also in the starburst galaxy NGC253 
(may be because of the bar shocks), 
but 50 times less in the starburst disk of M82 
(Garcia-Burillo et al 2001; Usero et al 2007).

\subsection{Outflows and chimneys}

Two distinct features are observed in the
 disk-halo interface in M82.
The associated large-scale shocks are traced by
SiO emission, extending out of the galaxy plane,
in a chimney (essentially to the North, see Fig 1)
and a supershell (mainly to the South).

In the SiO chimney, the molecular
filament extends out to 500pc in size
above the plane, the SiO abundance is estimated to
(2-4)x10$^{-10}$, and the corresponding H$_2$ mass at
M(H$_2$)$\sim$ 6x10$^6$ M$_\odot$
(Garcia-Burillo et al 2001).
The SiO supershell is the boundary of a cavity
filled by ionised gas around a supernova remnant, well
identified in the disk of M82. The supershell, of size 150pc, is still
closed towards the south (while it is broken towards the north).
It is associated to a supercluster of young stars. The
estimated H$_2$ mass is M(H$_2$) $\sim$ 1.6x10$^7$ M$_\odot$. 

\begin{figure}[ht]
\begin{center}
\includegraphics[width=8cm]{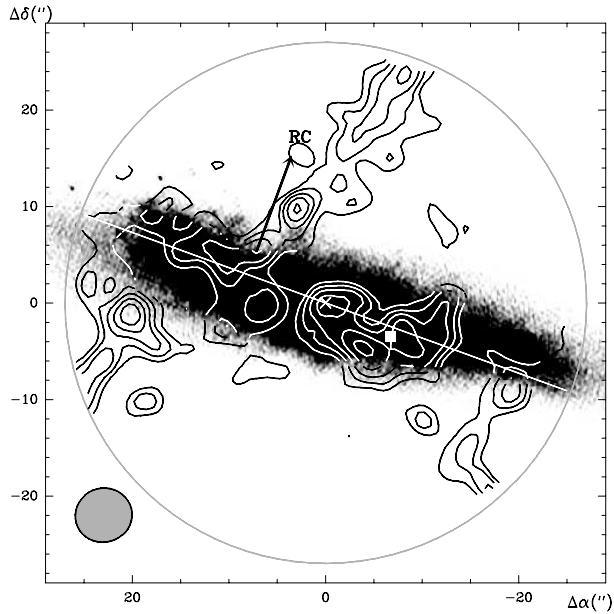}
\caption{The contours of SiO(v=0,J=2-1, at 87 GHz) emission
are superposed on the 4.8GHz radiocontinuum emission image 
in gray-scale, in the central region of M82.
The location of the radiocontinuum filament (RC) is highlighted
by an arrow, it delimits the chimney and the ejection of ionized gas,
towards the north. The supershell is to the south.
The outer circle represents the IRAM PdB primary beam (55''), while
the synthesized beam ($5.9''\times5.6''$) is shown at the bottom left.
The major axis of the galaxy is traced by the white line.
From Garcia-Burillo et al (2001).}
\label{fig1}
\end{center}
\end{figure}

\subsection{Shocks and bars}

The intense SiO emission in NGC253 could appear like a surprise,
since there is no AGN nor large chimneys and associated shocks,
although there exists a hot galactic wind entraining some molecular gas
(Garcia-Burillo et al 2000).
However, it seems that large-scale shocks due to the bar,
more precisely the crowding of clouds in resonant structures
(inner Lindblad resonances mainly),
enhance cloud-cloud collisions and heat the dust/gas to
produce the SiO signature.  Most of the SiO emission comes from 
a 600 x 250pc circumnuclear disk (CND) with a double ringed structure. The inner ring, 
of radius 60pc,  hosts the nuclear starburst; the outer pseudo-ring opens out 
as a spiral-like arc, in response  to the barred potential. 
The SiO abundance in the CND ($\sim$ 10$^{-9}$) is about one order of magnitude 
larger than what is expected in a PDR.

High excitation level NH$_3$ lines reveal that there is a large fraction
of warm (T=150K) gas in NGC~253, probably heated by C-schocks
in cloud-cloud collisions, while the gas is relatively cooler in M82
(Mauersberger et al 1996, 2003).

\section{PDR versus XDR}

\subsection{PDR chemistry in the starburst galaxy M82}

The M82 disk can be considered as a giant PDR of 600 pc in size
(Mao et al 2000; Garcia-Burillo et al 2002; Fuente et al 2005).
The strong UV field due to the compact starburst influences in depth the chemistry. 
Widespread HCO emission is detected with IRAM-PdB (Garcia-Burillo et al 2002),
and there are indications that PDR chemistry is propagating outward 
in the disk of the galaxy.
Global HCO abundances are comparable or even higher than PDR in the Milky Way,
  X(HCO)$\sim$ 4x10$^{-10}$ 
(Hollis \& Churchwell, 1983; Snyder et al 1985; Schilke et al 2001). 
  This high abundance of HCO in M82 puts the current PDR chemistry models into difficulties.
The formation of HCO is thought to be due to photo-destruction of
grains to produce H$_2$CO, followed by a photodissociation
of this molecule in the gas phase. But this process predicts an abundance
 X(HCO)$\sim$ 10$^{-11}$, at least 10 times lower.
In a similar way, PDR models cannot account for the high abundance
of CO$^+$, HOC$^+$ or c-C$_3$H$_2$. The latter abundance, as well as 
other small hydrocarbons, has been tentatively attributed to the desctruction
of PAH, but this dust grain chemistry is not yet fully included in the PDR
chemistry models.

\subsection{XDR chemistry in the Seyfert 2 galaxy NGC~1068}

X-rays are suspected to heavily influence molecular gas chemistry in the nuclear disks of AGN 
(Tacconi et al 1994, Maloney et al 1996).
While UV radiation is stopped by a column density of 10$^{21}$ cm$^{-2}$, hard X-rays
of energy 2-10kev can penetrate much further, up to 10$^{24}$ cm$^{-2}$. 
The first evidence of a highly different chemistry in extra-galactic XDR was
obtained from the high HCN/CO ratio (=1-10) in NGC~1068 (Tacconi et al 1994),
leading to a high HCN abundance. This anormal abundance could be due to 
oxygen depletion due to X-rays, yielding  a depletion of all oxygen-bearing molecules,
and in particular CO. Another possibility is that the X-rays enhance the ionisation,
itself enhancing the abundance of HCN, CN and molecular ions. Alternatively,
the X-rays could destroy small dust grains, and enhanced the frozen-in molecules,
and in particuler SiO. The multi-species observations of NGC~1068 by  
Usero et al (2004) have shown that oxygen-bearing molecules are not depleted,
ruling out the first hypothesis. SiO has been observed in the circum-nuclear disk
(CND) to be at least ten times higher than normal, supporting the X-induced
sputtering of dust grains (cf Fig 2).
One surprising result is the high HOC$^+$/HCO$^+$ ratio in the CND.
This supports also the scenario of a high ionisation towards the nucleus,
and confirms that the central part of NGC~1068 can be considered as a giant XDR.

\begin{figure}[ht]
\begin{center}
\includegraphics[width=13cm]{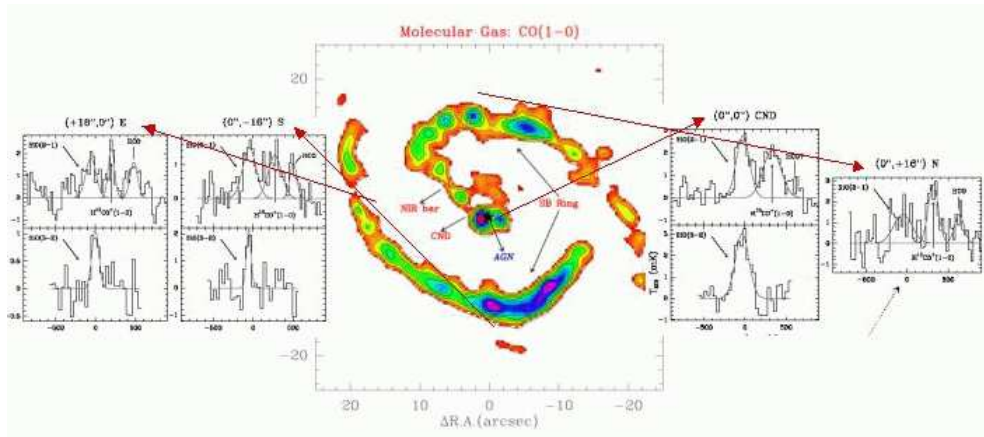}
\caption{Emission spectra of the 2--1 and 3--2 lines of SiO detected in the
inner 3~kpc of NGC~1068. The regions observed are indicated on the
CO(1--0) integrated intensity map of Schinnerer et al. (2000).
The central offset coincides with the position of the AGN, 
 while offsets at North, South and East probe the SiO emission over the
starburst ring.  From Usero et al (2004).}
\label{fig2}
\end{center}
\end{figure}

As for NGC~1068, NGC~1097 has an active nucleus, a Seyfert 1.
 HCN is enhanced in the center, and has been
derived to be a tracer of AGN (Kohno et al 2003).
While the HCN/CO emission ratio is equal to 0.1 in the nuclear
star-forming ring, it is 0.35 in the nucleus.

\subsection{Dynamics and chemistry in NUGA: NUclei of GAlaxies}

Observations at high spatial resolution with the IRAM
interferometer have been done towards a dozen nearby galaxies,
with an active nucleus (NUGA).
The CO gas morphology has been used to infer the gravitational 
torques due to the bar on the gas, and derive the efficiency of AGN fueling
(Garcia-Burillo et al 2005).
Statistically, the phase of fueling is not encountered frequently,
suggesting that the fueling is a short transient phase.

Some galaxies have also been observed in multi-species,
in order to test the influence of the AGN or starburst on
the chemistry (e.g. Krips et al 2007).
HCN(1-0), (2-1), (3-2), and HCO$^+$(1-0), (3-2)
have been observed at IRAM and also the sub-millimeter array (SMA).
It is frequent that both an AGN and a starburst are simultaneously
present in the galaxy centers, the starburst being a nuclear ring
encircling the nucleus. Spatial resolution is therefore required
to separate them.

Differences are indeed found in the line ratios between AGN and 
starbursts. There is a higher HCN emission, a lower HCO$^+$ emission and 
higher temperatures in AGN than in the starburst regions.
Similarly to NGC~1068, this might result from a higher HCN
 abundance in the centre due to an X-ray dominated gas chemistry,
 but a higher gas density/temperature or additional non-collisional
 excitation of HCN cannot be entirely ruled out. 
Indeed, a new result comes from the J=2-1, J=3-2 lines
of HCN and HCO$^+$. The
HCN/CO ratio decreases with J in AGN, while it 
remains constant in starbursts.

Low HCO$^+$ emission in AGN suggests that IR pumping is not important
 (however it might be important for HNC in ULIRGs,
     Guelin et al 2007, Aalto et al 2007).

The excitation ratios HCN(2-1)/HCN(1-0), or HCN(3-2)/HCN(1-0)
and HCO$^+$(3-2)/HCO$^+$(1-0) are found to be clear discriminators between
AGN and starbusrts, the latter being more excited
(Krips et al 2007, in prep). The excitation could come
from a higher density and/or higher temperature.
A modelisation through LVG models confirms that there
is relatively more dense gas in star-forming regions.
The H$_2$ density in regions around the AGN is then relatively low,
which may explain the low HCO$^+$/HCN in those XDR.

\begin{figure}[ht]
\begin{center}
\includegraphics[width=11cm]{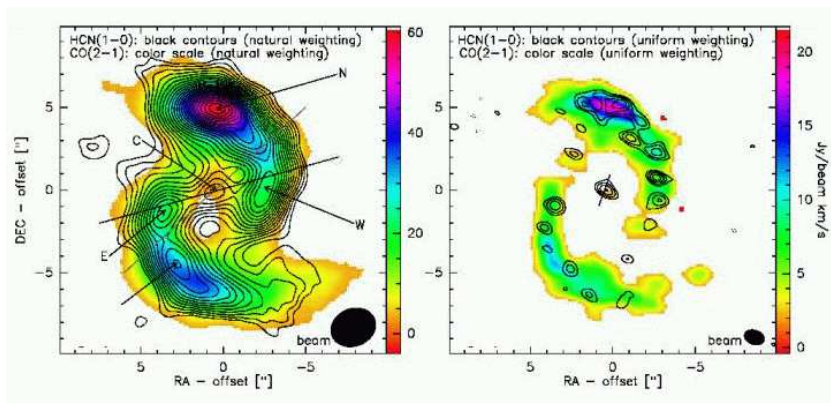}
\caption{Integrated HCN(1--0) emission ({\it black contours}) overlaid
on CO(2--1) ({\it color scale}) in natural ({\it left}) and
uniform weighting ({\it right}); the CO emission
has been smoothed to the same spatial resolution as the HCN.
The black line ({\it left})
indicates the major axis of the bar (PA=100$^\circ$) and the grey line
the major axis of the galaxy (PA=130$^\circ$). 
From Krips et al (2007).}
\label{fig3}
\end{center}
\end{figure}

A prototypical example of
the separation between AGN and starburst is the galaxy NGC~6951
(Fig 3). It is a barred spiral, with a ring of CO emission at
the inner Lindblad resonance.
This nuclear ring (400pc radius) is a clumpy starburst, while the
central component is AGN influenced.
There appears to be a molecular torus of 50pc in size,
which could be the obscuring component required by
the AGN unification model for this Seyfert 2.
The HCN/CO ratio is about 0.02 in the nuclear ring,
while it is  $>$ 0.4  in the torus
(Krips et al 2007).

\subsection{PDR and XDR models}

The differences between starburst and AGN chemistry and physical
conditions have been predicted by several models, with sometimes
non converging results  (e.g. Maloney et al 1996; Meijerink \& Spaans 2005).
Starbursts create regions dominated by UV photons (6-13.6ev)
from O \& B stars, e.g. PDR: they illuminate the 
surface of clouds, in the nuclear regions on scales $<$ 1kpc,
where higher HCO$^+$ is expected (from supernovae and associated cosmic rays).
Around AGN, molecular clouds are X-ray irradiated (1-100kev); X-ray 
penetrate more deeply into clouds of
the circum-nuclear disk CND. The radiation has then a volume effect, instead of a surface
effect in PDR.

The higher HCN/CO ratio in AGN, due to the XDR,
is indeed a prediction of models, but only for high (column) density gas, with columns 
in excess of 10$^{23}$ cm$^{-2}$ and densities larger than 10$^4$ cm$^{-3}$.
Otherwise, the HCN/CO ratio is lower in XDR than in PDR.

In recent models (e.g. Meijerink et al 2007) there is no
oxygen depletion in AGN, and CO emission is even enhanced with respect to PDR,
since the volume in which CO is excited is much wider.
Contrary to the observations, models predict a low HCN/HCO$^+$ ratio
at high density in the XDR, i.e. when the HCN/CO ratio is high.
The contrary is expected to be true at low density.
According to density and irradiation 
strength, it should be possible to find 
HCN/HCO$^+$ enhanced in XDR.
On the other hand, CO(1-0)/H$_2$ is predited to be lower in XDR,
since the CO molecule is warmer, and high-J levels are excited,
which could also explain the HCN/CO(1-0) ratio.

\subsection{Is the HCN/CO ratio a star formation indicator?}

It is now well established that, while the
HCN/CO ratio is about 0.1 or less in normal galaxies,
it is enhanced to values larger than 0.3 and even up to 1
towards the nuclei of AGN (Kohno et al 1999, Usero et al 2004).

There is however another observation of boosted HCN emission
towards starburst galaxies, and in particular 
ultraluminous in the infrared: there is in particular a
good correlation between HCN and IR due to star formation
(Gao \& Solomon 2004). This correlation is not likely due
to infrared pumping through UV/X-ray heated dust,
since there is no correlation between X-ray and MIR.

The far infrared luminosity is thought to be one 
of the best tracers of star formation, and the fact
that it is better correlated to HCN than to CO is
interpreted as a correlation with dense gas (HCN
being here a dense gas tracer essentially).

While the relation between the star formation 
rate and CO is non-linear, it is 
linear with the HCN(1-0) emission.
The SFR is directly proportional
to the dense gas, and it is to the latter that
 the Schmidt law is related.

We are therefore left with a degeneracy, towards
the central parts of ULIRGs: either the HCN is a
tracer of dense gas, and therefore of star formation
(Gao \& Solomon 2004), or the importance of the AGN
is highly increasing at high infrared luminosity
(as has been already observed, Tran et al 2001, Veilleux et al 2002),
and most of the HCN emission could come from abundance
effects due to an XDR. 

HCO$^+$, which is also a dense gas tracer, might
help to raise the degeneracy. 
 The problem is complex however, since not
only density, but also temperature, excitation
and physical effects play a role, and 
in starbursting galaxies, the nucleus activity is often symbiotic
to the star-forming region.
According to the low HCO$^+$/CO ratio observed in ULIRGs by
Gracia-Carpio et al (2006), cf Fig 4, and the
correlation found between
the HCN/HCO$^+$ ratio and the IR luminosity of the galaxy (LIR),
the HCN enhancement in ULIRGS could be due only to the XDR
chemistry, and not to the dense gas and star formation.

\begin{figure}[ht]
\begin{center}
\includegraphics[width=6cm]{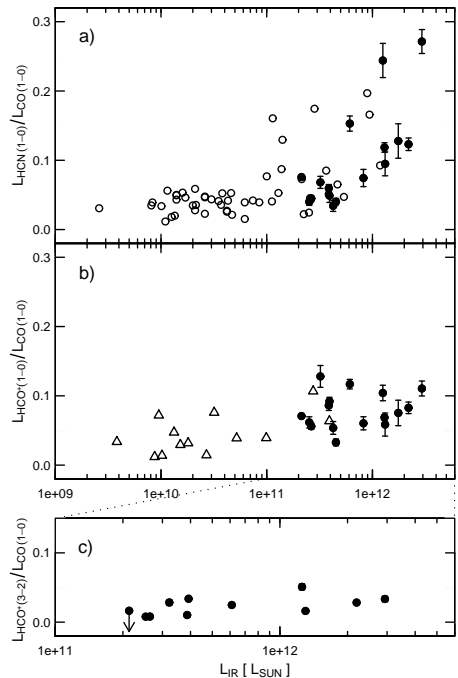}
\caption{Variation of the HCN(1--0)/CO(1--0) ({\bf top}) and HCO$^+$(1--0)/CO(1--0) 
({\bf middle}) luminosity ratios from
normal galaxies ($L_{\rm{IR}} < 10^{11}\,L_{\odot}$) to LIRGs and ULIRGs 
($L_{\rm{IR}} > 10^{11}\,L_{\odot}$). The HCO$^+$(3--2)
/CO(1--0) ratio ({\bf bottom}) is about constant for LIRGs and ULIRGs. 
From Gracia-Carpio et al (2006).}
\label{fig4}
\end{center}
\end{figure}

\section{Conclusions}

The abundant observations of many different chemical and
density tracers in external galaxies have shown that
chemistry is indeed coupled to dynamics, via shocks, bars, 
galactic outflows.

Chemical and density tracers could help to discriminate 
between starburst and AGN as the source of molecular emission.
In particular, photodissociation regions (PDR)
and X-ray dominated regions (XDR) bear different 
diagnostics in HCN, CO and HCO$^+$ ratios.

The fact that the HCN abundance is enhanced in XDR, 
brings a degeneracy in the HCN/CO tracer of dense gas 
and of star formation rate. In nuclei of galaxies, 
other tracers should be also searched for in parallel.
In all cases, enhanced spatial resolution is required to raise
the degeneracy.


\end{document}